\pdfoutput=1  
\def\nk{n_\mathrm{K}}
\def\acap{\\ \nonumber \\}

\def\rfr#1{Equation\,(\ref{#1})}
\def\rfrs#1#2{Equations\,(\ref{#1})--(\ref{#2})}

\def\Rfrs#1#2{Equations\,(\ref{#1})--(\ref{#2})}

\def\eqi{\begin{equation}}
\def\eqf{\end{equation}}
\def\rp#1#2{\frac{#1}{#2}}
\def\lb#1{\label{#1}}

\def\ton#1{\left(#1\right)}
\def\qua#1{\left[#1\right]}
\def\grf#1{\left\{#1\right\}}

\RequirePackage[2020-02-02]{latexrelease}
\documentclass{aastex631}
\usepackage{morefloats}
\usepackage[title]{appendix}
\usepackage{textcomp}
\usepackage{booktabs}
\usepackage{multirow}
\usepackage{rotating,tabularx}
\usepackage{float}
\usepackage{enumerate}
\usepackage{rotating}
\usepackage[polutonikogreek,english]{babel}
\usepackage{amsmath,starfont,textgreek,w-greek}
\usepackage[flushleft]{threeparttable}
\usepackage{amsthm}
\usepackage{bigints}
\usepackage{amscd}
\usepackage[mathlines]{lineno}
\usepackage{amssymb,dsfont}
\usepackage{graphicx,epsfig}
\usepackage{txfonts}
\usepackage{xr-hyper}
\usepackage{hyperref}
\usepackage[utf8]{inputenc}
\usepackage{newunicodechar,graphicx}

\usepackage[nointegrals]{wasysym}
\usepackage[caption=false]{subfig}

\allowdisplaybreaks

\makeatletter
\DeclareRobustCommand\ref{%
    \@ifstar\@refstar\T@ref
  }%
  \DeclareRobustCommand\pageref{%
    \@ifstar\@pagerefstar\T@pageref
  }%
 \makeatother

\begin{document}

\title{Using the Difference of the Inclinations of a Pair of Counter-Orbiting Satellites to Measure the Lense--Thirring Effect}

\shortauthors{L. Iorio}

\author[0000-0003-4949-2694]{Lorenzo Iorio}
\affiliation{Ministero dell' Istruzione e del Merito. Viale Unit\`{a} di Italia 68, I-70125, Bari (BA),
Italy}

\email{lorenzo.iorio@libero.it}

\begin{abstract}
Let two test particles A and B revolving about a spinning primary along \textrm{ideally} identical orbits in \textrm{opposite} directions be considered. From the \textrm{general} expressions of the precessions of the orbital \textrm{inclination} induced by the post--Newtonian gravitomagnetic and Newtonian quadrupolar fields of the central object, it turns out that the \textrm{Lense--Thirring} inclination rates  of  A and B are \textrm{equal} and \textrm{opposite}, while the \textrm{Newtonian} ones due to the primary's oblateness are \textrm{identical}. Thus, the \textrm{difference} of the inclination shifts of the two orbiters would allow, in principle, to \textrm{cancel out} the \textrm{classical} effects by \textrm{enhancing} the general relativistic ones. The conditions affecting the orbital configurations that must be satisfied for this to occur and possible observable consequences in the field of Earth  are investigated. In particular, a scenario involving two spacecraft in polar orbits, branded POLAr RElativity Satellites (POLARES) and reminiscent of an earlier proposal by Van Patten and Everitt in the mid--1970s, is considered. A comparison with the ongoing experiment with the LAser GEOdynamics Satellite (LAGEOS) and LAser RElativity Satellite (LARES) 2 is made.
\end{abstract}


\keywords{classical general relativity; experimental studies of gravity;  experimental tests of gravitational theories; satellite orbits; harmonics of the gravity potential field}
%
%
%
\section{Introduction}
To the first post--Newtonian (1pN) order, the General Theory of Relativity (GTR) predicts, among other things, that the orbital motion of a test particle freely orbiting a massive primary undergoes certain long--term, cumulative perturbations due to the gravitomagnetic field of the central object caused by its spin angular momentum. They are called the  Lense--Thirring (LT) effect \cite{1918PhyZ...19..156L,1984GReGr..16..711M}, although recent historical studies \cite{2007GReGr..39.1735P,2008mgm..conf.2456P,Pfister2014} pointed out that it would be more correct to rename it as Einstein--Thirring--Lense effect. Basically, it consists of variations of the orientation of both the orbital plane and of the orbit  within the orbital plane itself which manifest themselves cumulatively revolution after revolution. Instead, the shape and the size of the path are left unaffected along with the time of passage at the pericentre. The LT effect is quite small in ordinary weak--field and slow--motion scenarios like the surroundings of, say, the Earth or the Sun. Suffice it to say that the perihelion of Mercury, whose orbital period amounts to about 88 days, is shifted by the solar angular momentum by just 2 milliarcseconds per century (mas cty$^{-1}$). Moreover, the orbital plane of the LAser GEOdynamics Satellite (LAGEOS) \cite{1985JGR....90.9217C}, revolving about the Earth in less than 4 hours, precesses at a rate as little as a few tens of milliarcseconds per year (mas yr$^{-1}$) due to the terrestrial gravitomagnetic field. On the other hand, such a general relativistic feature of motion should play a decisive role in the intricate dynamics of accreting matter close to Kerr black holes \cite{1975ApJ...195L..65B}. For example, it should drive the relativistic jets emanating from the surroundings of the supermassive black holes lurking in the active nuclei of radio galaxies \cite{1978Natur.275..516R,1984ARA&A..22..471R}. Furthermore, after an accretion disk is formed around a supermassive black hole, initially with a strong misalignment with respect to the spin of the latter, as a consequence of a tidal disruption event of a nearby passing star, the LT effect causes the former to precess at early times before it finally aligns with the hole's equatorial plane, ending the precession  \cite{2012PhRvL.108f1302S,2016MNRAS.455.1946F,2024Natur.630..325P}. Finally, it is believed to cause quasi periodic oscillations in the X--ray flux of accreting compact objects \cite{1998ApJ...492L..59S,2009SSRv..148..105S}. In such extreme natural laboratories, the expected magnitude of the aforementioned LT--driven effects is large, thus not posing a challenge to their detection. Rather, it is the interpretation of these phenomena that is difficult because of lot competing effects whose physical mechanisms are not yet understood with enough accuracy \cite{2009SSRv..148..105S}. As a consequence, it is of the utmost importance to have accurate and reliable tests of the LT effect  performed in better known environments  in order to extrapolate its validity also to the aforementioned strong field scenarios.

A wealth of gravitomagnetic effects other than the LT one exist to all scales \cite{1977PhRvD..15.2047B,1986SvPhU..29..215D,2002EL.....60..167T,2002NCimB.117..743R,2004GReGr..36.2223S,2009SSRv..148...37S}.
Among them, the Pugh--Schiff \cite{Pugh59,Schiff60} precession of the spin axes of four gyroscopes carried onboard a drag--free spacecraft orbiting the spinning Earth was successfully measured with a $19\%$ accuracy \cite{2011PhRvL.106v1101E,2015CQGra..32v4001E} by the dedicated Gravity Probe B (GP-B) mission \cite{Varenna74,2001LNP...562...52E}. To date, it still remains the only undisputed test of a gravitomagnetic effect in the existing peer--reviewed literature.

Returning to the LT effect, tests of it performed in astronomical scenarios in the solar system are quite rare. At present, there are reports in the literature of experiments made with Mercury in the Sun's field \cite{2015IAUGA..2227771P,Pav2024,RussiLT} and with the Juno probe \cite{2017SSRv..213....5B} around Jupiter \cite{2011AGUFM.P41B1620F,2024ApJ...971..145D}. Although none of them is in disagreement with the predictions of GTR, the reported uncertainties and the correlations with other estimated parameters are large enough to make the obtained results inconclusive.

As far as the terrestrial field is concerned, attempts have been underway for almost 30 years \cite{1996NCimA.109..575C} to measure the LT orbital precessions using some geodetic satellites \cite{2019JGeod..93.2181P} tracked with the Satellite Laser Ranging (SLR) technique \cite{SLR11}; for reviews, see, e.g., \cite{2011Ap&SS.331..351I,2013NuPhS.243..180C,2013CEJPh..11..531R}, and references therein.
Although the pericenter  is also impacted by the gravitomagnetic field, for 20 years now one has focused on the nodes of some satellites of the LAGEOS family \cite{2013NuPhS.243..180C,2019Univ....5..141L,2020Univ....6..139L}. Such a choice is due to the fact that the node of a satellite is much less severely disturbed than the perigee by the competing non--gravitational accelerations \cite{Nobilibook87,2001P&SS...49..447L,2002P&SS...50.1067L,2004CeMDA..88..269L,2004P&SS...52..699L}. In 1976, Van Patten and Everitt \cite{1976CeMec..13..429V,1976PhRvL..36..629V} proposed to look at the sum of the nodes of a pair of low--altitude, drag--free spacecraft moving in opposite directions along ideally identical circular orbits passing through the Earth's poles. Indeed, while the LT precessions add up, the nominally much larger competing Newtonian node shifts due to the Earth's quadrupole mass moment, which would act as a major source of systematic bias, cancel out, in principle, for such an orbital configuration. An essentially equivalent version of such an idea, which its promoters intended would have allowed a  $\simeq 1\%$ measurement of the LT effect, was put forth 10 years later by Ciufolini \cite{1986PhRvL..56..278C} who proposed to use a pair of passive SLR satellites following ideally identical non--polar orbits whose inclinations to the Earth's equator are displaced by 180$^\circ$. Indeed, such a scenario is conceptually equivalent to the one by Van Patten and Everitt, apart from the technical details pertaining the tracking method and the mechanism of compensation of the non--gravitational perturbations, since, also in this case, the classical node precessions cancel out while the LT ones add up. The idea by Ciufolini \cite{1986PhRvL..56..278C} came to fruition in the last years with the launch of the LAser RElativity Satellite (LARES) 2 \cite{2023EPJC...83...87C} in July 2022 joining LAGEOS, which had already been in orbit for almost 50 years. Ciufolini and coworkers \cite{2023EPJC...83...87C} claimed that it would be possible to perform a LT test with such satellites accurate to $\simeq 0.2\%$. Unfortunately, their actual orbital configurations are different just enough to not allow the cancellation of the classical precessions to a good enough level \cite{2023Univ....9..211I}.

Here, the proposal of using a pair of counter--revolving satellites in polar orbits, collectively dubbed POLAr RElativity Satellites (POLARES), is reexamined by showing that it would ideally be possible to use not only the sum of their nodes but also the difference of their inclinations to extract the LT effect reducing the biasing impact of the Earth's oblateness to an acceptable level. In principle, should the orbits be sufficiently elliptical, also the difference of the perigees could be adopted \cite{IorioPLA03}. However, this would likely force to use expensive drag--free technologies to counterbalance the non--gravitational perturbations the perigees of geodetic satellites are particularly sensitive to.

The following physical and orbital parameters will be used in the rest of the paper. Among the constants of Nature, $G$ is the Newtonian constant of gravitation, and $c$ is the speed of light in vacuum. As far as the key physical parameters of the Earth are concerned,  $\upmu:=GM$ is the standard gravitational parameter given by the product of $G$ times the mass $M$, $\boldsymbol{J}=J\boldsymbol{\hat{k}}$ is the spin angular momentum,  $\boldsymbol{\hat{k}}$ is the spin axis, 
$R$ is the  mean equatorial radius,  $J_2$ is the first even zonal harmonic coefficient of degree $\ell =2$ and order $m=0$ of the multipolar expansion of the geopotential accounting for deviations from spherical symmetry, and $\rho_\mathrm{atm}$ is the atmospheric density ay some height. The relevant orbital parameters characterizing the satellite's motion referred to some Earth--centered inertial  (ECI) reference frame are the semimajor axis $a$, the eccentricity $e$, the semilatus rectum $p:=a\ton{1 - e}$, the inclination of the orbital plane $I$, and the longitude of the ascending node $\Omega$. Furthermore, $\nk:=\sqrt{\upmu/a^3}$ is the Keplerian mean motion.
%
%
%
%

The paper is organized as follows. In Section \ref{sec:gen}, the general expressions for the classical and relativistic rates of change of the inclination and the nodes, averaged over one orbital revolution and valid for an arbitrary orientation of the primary's spin axis, are reviewed. Their consequences for counter--revolving satellites in polar orbits are discussed, with particular emphasis on the difference of the inclinations. The impact of the orbital injection errors on the difference of the inclinations and the sum of then nodes is the subject of Section \ref{orb_err}. A comparison with LAGEOS and LARES 2 is made in Section \ref{cazzate}. Section \ref{fine} summarizes the findings and offers conclusions.
\section{The general expressions for the LT and $J_2$ precessions of the orbital plane and their consequences}\lb{sec:gen}
The spin axis of a celestial body in the solar system is usually parameterized as
\eqi
\boldsymbol{\hat{k}} = \grf{\cos\alpha\cos\delta,\sin\alpha\cos\delta,\sin\delta}\lb{kvers}
\eqf
in terms of the right ascension (R.A.) $\alpha$ and declination (decl.) $\delta$ of its north pole of rotation referred to the Earth's Mean Equator and Mean Equinox (MEME) at 12:00 Terrestrial Time on 1 January 2000 (J2000.0). It should be noted that, in general, $\alpha,\delta$ are time--dependent because of possible gravitational pulls exerted by other more or less distant bodies; in the case of Earth, they induce, e.g., the lunisolar precession and nutation \cite{PrecNut}.

In view of the following developments, it is useful to introduce the unit vectors $\boldsymbol{\hat{l}},\boldsymbol{\hat{m}},\boldsymbol{\hat{h}}$  defined as
\begin{align}
\boldsymbol{\hat{l}} \lb{elle} & = \grf{\cos\Omega,\sin\Omega,0},\acap
\boldsymbol{\hat{m}} \lb{emme} & = \grf{-\cos I \sin\Omega,
\cos I \cos\Omega, \sin I},\acap
\boldsymbol{\hat{h}} \lb{acca} & = \grf{\sin I \sin\Omega, -\sin I \cos\Omega, \cos I}
\end{align}
in such a way that $\boldsymbol{\hat{l}}\boldsymbol\times\boldsymbol{\hat{m}} = \boldsymbol{\hat{h}}$ holds. The unit vector $\boldsymbol{\hat{l}}$ is directed along the line of nodes towards the ascending node, while $\boldsymbol{\hat{h}}$ is aligned with the satellite's orbital angular momentum.

The LT and Newtonian precessions of the inclination $I$ and node $\Omega$, valid for an arbitrary orientation of the primary's spin axis $\boldsymbol{\hat{k}}$ with respect to the inertial system adopted, are
\begin{align}
\dot I^\mathrm{LT} \lb{dotILT}& =\rp{2GJ}{c^2 a^3\ton{1 - e^2}^{3/2}}\boldsymbol{\hat{k}}\boldsymbol\cdot\boldsymbol{\hat{l}} = \rp{2GJ\cos\delta\cos\ton{\alpha-\Omega}}{c^2 a^3\ton{1 - e^2}^{3/2}},\acap
\dot I^{J_2} \lb{dotIJ2} & = -\rp{3}{2}\nk J_2\ton{\rp{R}{p}}^2\ton{\boldsymbol{\hat{k}}\boldsymbol\cdot\boldsymbol{\hat{l}}}\ton{\boldsymbol{\hat{k}}\boldsymbol\cdot\boldsymbol{\hat{h}}} =  \rp{3}{2}\nk J_2\ton{\rp{R}{p}}^2 \cos\delta \cos\ton{\alpha - \Omega} \qua{-\cos I \sin\delta + \sin I\cos\delta\sin\ton{\alpha - \Omega}}, \acap
\dot \Omega^\mathrm{LT} \lb{dotOLT}& =\rp{2GJ\csc I}{c^2 a^3\ton{1 - e^2}^{3/2}}\boldsymbol{\hat{k}}\boldsymbol\cdot\boldsymbol{\hat{m}} = \rp{2GJ\qua{\sin\delta  + \cos\delta  \cot I \sin\ton{\alpha - \Omega}}}{c^2 a^3\ton{1 - e^2}^{3/2}},\acap
\dot \Omega^{J_2} \nonumber & = -\rp{3}{2}\nk J_2\ton{\rp{R}{p}}^2\csc I\ton{\boldsymbol{\hat{k}}\boldsymbol\cdot\boldsymbol{\hat{m}}}\ton{\boldsymbol{\hat{k}}\boldsymbol\cdot\boldsymbol{\hat{h}}} = \acap
\lb{dotOJ2} &= \rp{3}{2}\nk J_2\ton{\rp{R}{p}}^2 \sin I\qua{-\cot I \sin\delta  +
 \cos\delta  \sin\ton{\alpha - \Omega}}\qua{\sin\delta  + \cot I\cos\delta \sin\ton{\alpha - \Omega}}.
\end{align}
Interestingly, the LT shift of the inclination given by \rfr{dotILT} does not depend on the inclination itself.

From \rfrs{dotILT}{dotIJ2} it turns out that, if the primary's spin axis is aligned with the reference $z$ axis, corresponding to
\eqi
\delta=90^\circ,\lb{d90}
\eqf
both the LT and the classical rates of change of $I$ vanish, contrary to the node shifts which reduce to the well known secular precessions
\begin{align}
\dot\Omega^\mathrm{LT} \lb{OLT}& = \rp{2GJ}{c^2 a^3\ton{1-e^2}^{3/2}},\acap
\dot\Omega^{J_2} \lb{OJ2}& = -\rp{3}{2}\nk J_2\ton{\rp{R}{p}}^2\cos I.
\end{align}
widely used in the literature, as per \rfrs{dotOLT}{dotOJ2}. The entire body of published works on the SLR--based LT tests, including \cite{1986PhRvL..56..278C}, rely upon  \rfrs{OLT}{OJ2}, while the inclination has never been considered so far in this context.

If $\boldsymbol{\hat{k}}$ is generally not aligned with the reference $z$ axis, the situation goes as follows.
If A and B denote two satellites moving along orbits with ideally identical shapes and sizes, the condition to be met for them to move along opposite directions is
\begin{align}
I_\mathrm{B} \lb{IAB} & =180^\circ -I_\mathrm{A},\acap
\Omega_\mathrm{B} \lb{OAB} & = \Omega_\mathrm{A} + 180^\circ.
\end{align}
Indeed, from \rfr{acca} and \rfrs{IAB}{OAB}, it turns out just
\eqi
{\boldsymbol{\hat{h}}}_\mathrm{B} = - {\boldsymbol{\hat{h}}}_\mathrm{A}.\lb{hAB}
\eqf
Moreover, it is also
\begin{align}
{\boldsymbol{\hat{l}}}_\mathrm{B} \lb{lAB}&= - {\boldsymbol{\hat{l}}}_\mathrm{A},\acap
{\boldsymbol{\hat{m}}}_\mathrm{B} \lb{mAB}&= {\boldsymbol{\hat{m}}}_\mathrm{A},
\end{align}
so that
\eqi
{\boldsymbol{\hat{l}}}_\mathrm{B}\boldsymbol\times{\boldsymbol{\hat{m}}}_\mathrm{B} = {\boldsymbol{\hat{m}}}_\mathrm{A}\boldsymbol\times{\boldsymbol{\hat{l}}}_\mathrm{A} = -{\boldsymbol{\hat{h}}}_\mathrm{A} = {\boldsymbol{\hat{h}}}_\mathrm{B}.
\eqf

From \rfrs{dotILT}{dotIJ2} and \rfrs{hAB}{mAB}, it follows that, for a given orientation of $\boldsymbol{\hat{k}}$, the LT inclination rates are equal and opposite, while the classical ones are identical.
Thus, in principle, one can look at the difference of the inclination rates of two counter--orbiting satellites
\eqi
\dot I^\mathrm{A} - \dot I^\mathrm{B}\lb{DI}
\eqf
since the LT effect would be enhanced, while the competing classical shifts would exactly cancel out.

Instead, the opposite holds for the node shifts: the LT rates are identical, while the $J_2$--driven ones are opposite. Incidentally, this proves that, independently of the actual value of the inclination\footnote{The actual value of $I$ did not enter the above reasonings.} of the orbital planes and of the orientation of the primary's spin axis, the counter--orbiting scenario is conceptually equivalent to the LAGEOS--LARES 2 one by Ciufolini \cite{1986PhRvL..56..278C}, relying upon \rfrs{OLT}{OJ2},  in the sense that also for a pair of counter--revolving satellites the sum of the LT node rates add up, while the Newtonian ones cancel out. However, it should be stressed that the orbital geometry proposed by Ciufolini \cite{1986PhRvL..56..278C} is not supplemented by any condition on the satellites' nodes. Thus, for a general orientation of $\boldsymbol{\hat{k}}$, the sole condition on the inclination given by \rfr{IAB} does not allow to cancel out the classical node precessions by summing them, not even in the ideal case of identical semimajor axes and eccentricities. Indeed, by imposing only \rfr{IAB}, one has
\eqi
\ton{\boldsymbol{\hat{k}}\boldsymbol\cdot{\boldsymbol{\hat{m}}}_\mathrm{A}}\ton{\boldsymbol{\hat{k}}\boldsymbol\cdot{\boldsymbol{\hat{h}}}_\mathrm{A}} +
\ton{\boldsymbol{\hat{k}}\boldsymbol\cdot{\boldsymbol{\hat{m}}}_\mathrm{B}}\ton{\boldsymbol{\hat{k}}\boldsymbol\cdot{\boldsymbol{\hat{h}}}_\mathrm{B}} \nonumber =\cos\delta \qua{\sin\ton{\alpha - \Omega_\mathrm{A}} + \sin\ton{\alpha - \Omega_\mathrm{B}}} \qua{\cos 2 I_\mathrm{A} \sin\delta +
\cos\delta \cos\ton{\alpha - \rp{\Sigma\Omega}{2}} \sin 2 I_\mathrm{A} \sin\rp{\Delta\Omega}{2}},\lb{intuculo}
\eqf
where
\begin{align}
\Sigma\Omega & :=\Omega_\mathrm{A} + \Omega_\mathrm{B}, \acap
\Delta\Omega & :=\Omega_\mathrm{A} - \Omega_\mathrm{B}.
\end{align}

If condition
\begin{align}
I \lb{pol1} & = 90^\circ,\acap
\Omega \lb{pol2} & = \alpha,
\end{align}
implying that the orbit is polar since \rfrs{pol1}{pol2} yield
\eqi
\boldsymbol{\hat{k}}\boldsymbol\cdot\boldsymbol{\hat{h}}=0,
\eqf
is also imposed, then the LT inclination rate of \rfr{dotILT}, which is independent of $I$, does not vanish, while the classical one does so, as per \rfr{dotIJ2}.

Thus, if \rfr{d90} does not hold, a pair of counter--orbiting satellites moving along identical polar orbits would allow, in principle, to measure the LT effect also using the difference of their inclinations as well as the sum of their nodes.

Does all this have any practical relevance in the case of a possible mission around the Earth? The answer is positive for the following reasons.
The ECI which is routinely used in satellite's data reductions is the Geocentric Celestial Reference System (GCRS) \cite{iers10}. It is essentially characterized by the MEME, being also dubbed as J2000 system.
More precisely, the orientation of GCRS coincides by default with that of the International Celestial Reference System (ICRS), as per the Recommendation 2 of the IAU 2006 Resolution B.2 by the International Astronomical Union (IAU) \cite{iers10}. In turn, the fundamental plane of ICRS is almost coincident with the Earth's mean equator at J2000.0, up to a constant offset, known as frame bias, as little as a few tens milliarcseconds \cite{2006A&A...450..855C}.
Thus, the reference $z$ axis of the ECI adopted is substantially aligned with the Earth's spin axis at J2000.0.
The data analyses of any future satellite--based mission aimed at measuring the LT effect will necessarily be carried out over a time span during which the terrestrial spin axis will not coincide with the J2000 one due to, e.g., the lunisolar precession. Thus, the general expressions of \rfrs{dotILT}{dotOJ2} are to be used implying, among other things,  that also the LT effect on the inclination can be looked at; according to \rfr{dotILT}, the later than the year 2000 the mission launches or data analysis begins, the greater the LT effect on inclination. By the way, the same considerations should be extended also to the current LAGEOS--LARES 2 experiment since the latter one was launched about two years after J2000.0 and their data analyses will continue for several years onwards.
\section{The impact of the unavoidable departures of the actual orbits from the ideal ones}\lb{orb_err}
The final orbital configurations of the satellites once launched would differ from their idealized counterparts because of the unavoidable orbit injection errors. Scope of this Section is investigating their impact on the level of cancellation of the classical perturbations due to the Earth's oblateness which can be actually achievable by taking the difference of the inclinations (Section \ref{sub:incli})  and the sum of the nodes (Section \ref{sub:nodi}) of the two POLARES.

To this aim, the equations for the rates of change of $I$ and $\Omega$, averaged over one orbital revolution, were simultaneously integrated with respect to time over a time span 10 years long by inserting \rfrs{dotILT}{dotOJ2} in their right--hand--sides in order to obtain time--dependent shifts $\Delta I\ton{t},\Delta\Omega\ton{t}$ for both satellites. Furthermore, also the secular trend and the annual harmonic variations of $J_2$,  as modeled in the Earth's gravity model ITSG-Grace2018,  retrievable at \url{http://doi.org/10.5880/ICGEM.2018.003}, were taken into account. Finally, the precessional motion of the Earth's spin axis was included according to \cite[pp.\,176-177]{2000Monte} as well. The start date was assumed to be, say, 35 years after J2000.0, corresponding to a hypothetical launch in the next ten years. In each integration, the initial values of the semimajor axes, the eccentricities, the nodes and the inclinations were modified from time to time by small quantities compared to their ideal counterparts in order to simulate orbit injection errors.
\subsection{The difference of the inclinations}
\begin{figure}
\centering
\begin{tabular}{c}
\includegraphics[width = 8 cm]{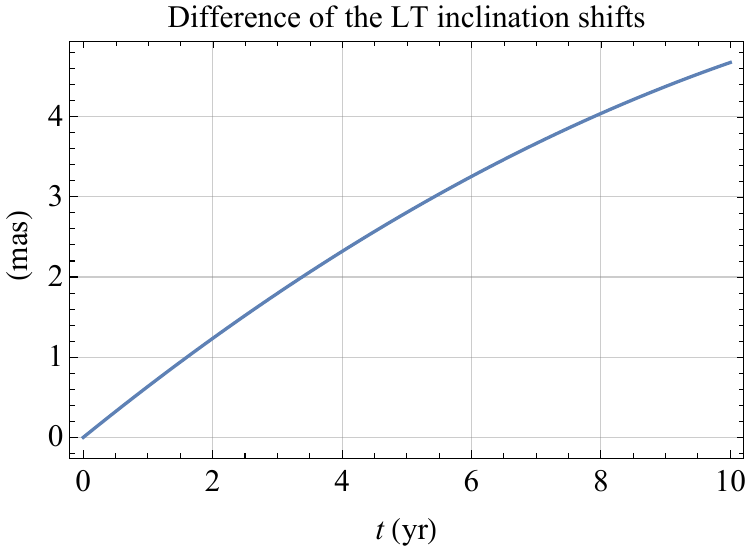}\\
\includegraphics[width = 8 cm]{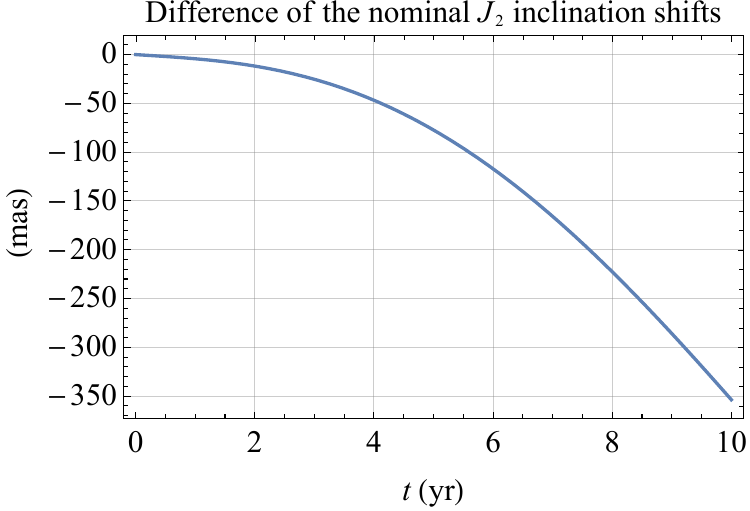}\\
\includegraphics[width = 8 cm]{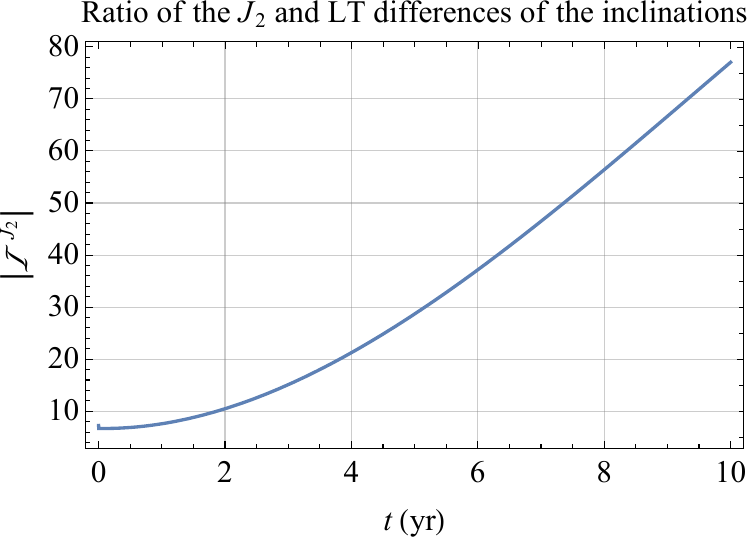}\\
\end{tabular}
\caption{Differences of the nominal  LT (upper panel) and $J_2$ (middle panel) shifts of the inclinations, in mas, of a pair of counter--orbiting satellites numerically integrated over 10 years. The temporal variations of both $\boldsymbol{\hat{k}}$ \cite[pp.\,176-177]{2000Monte} and $J_2$, modeled according to ITSG-Grace2018, were included as well. An initial epoch 35 years after J2000.0 was assumed. An orbital height of $2\,000$ km was adopted for both satellites up to an offset of 10 km. The initial values of the inclinations differ from \rfrs{IAB}{OAB} and \rfrs{pol1}{pol2} by 10 mas. The lower panel shows the plot of the absolute value of \rfr{DeltaIJ2LT}.
}\label{figure1}
\end{figure}

\lb{sub:incli}
Figure \ref{figure1}
shows the plots of the differences of the nominal integrated shifts of the inclinations induced by the LT effect and the Earth's oblateness obtained for an orbital height of $2\,000$ km for both spacecraft up to 4 km and almost circular orbits whose eccentricities differ by $0.00376$. \Rfrs{IAB}{OAB} and \rfrs{pol1}{pol2} were used for the initial values $I_0,\Omega_0$ of the inclinations and the nodes up to offsets of 10 mas in $I_0$ and 10 arcseconds in $\Omega_0$.
Furthermore, also the plot of the absolute value of
\eqi
\mathcal{I}^{J_2}\ton{t}:=\rp{\Delta I^{J_2}_\mathrm{A}\ton{t} - \Delta I^{J_2}_\mathrm{B}\ton{t}}{\Delta I^\mathrm{LT}_\mathrm{A}\ton{t} - \Delta I^\mathrm{LT}_\mathrm{B}\ton{t}}\lb{DeltaIJ2LT}
\eqf
is depicted. The ratio $\mathcal{I}^{J_2}$ is a measure of the nominal systematic bias induced by the Earth's quadrupole mass moment on the expected LT signal; the larger it is, the greater the indirect impact of the errors of the various parameters ($G, J_2, R, \upmu, J, a_\mathrm{A,B}, e_\mathrm{A,B}, I_\mathrm{A,B},\Omega_\mathrm{A,B},\ldots$) entering it.
It turns out that the expected LT signal is at the mas level. Instead, the $J_2$ one is just up to about 80 times larger. Such a feature is important since it allows to make the mismodeling in \rfr{DeltaIJ2LT} induced by the several sources of errors affecting it  negligible.
By varying the offsets in the orbital elements, it turns out that  differences in the values of the semimajor axes and the eccentricities as large as those of the existing LAGEOS and LARES 2 \cite{2023EPJC...83...87C} are well tolerated.  Discrepancies of the initial values of the nodes from their ideal values of \rfr{OAB} and \rfr{pol2}  up to  $\delta\Omega_0\simeq 0.1^\circ$ would not yield a dramatic increase of $\mathcal{I}^{J_2}$. As far as the inclinations are concerned,  departures  from the ideal values of \rfr{IAB} and \rfr{pol1}  up to $\delta I\simeq 100\,\mathrm{mas}$ would not  change the pattern of Figure\,\ref{figure1}.
\subsection{The sum of the nodes}
\begin{figure}
\centering
\begin{tabular}{c}
\includegraphics[width = 8 cm]{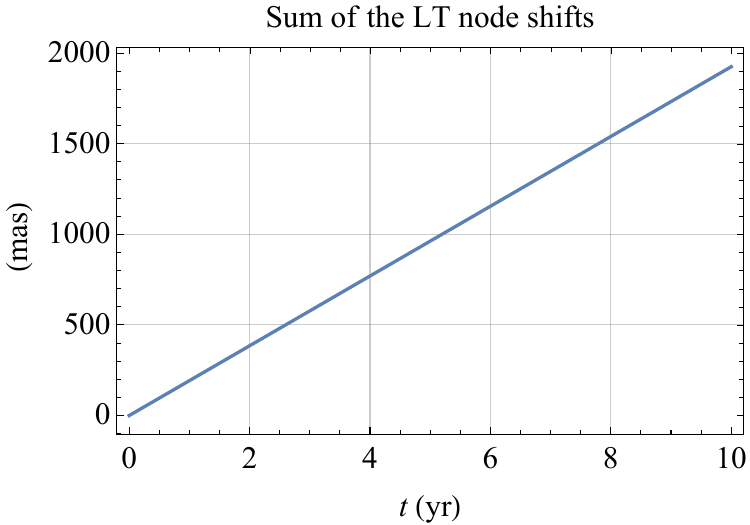}\\
\includegraphics[width = 8 cm]{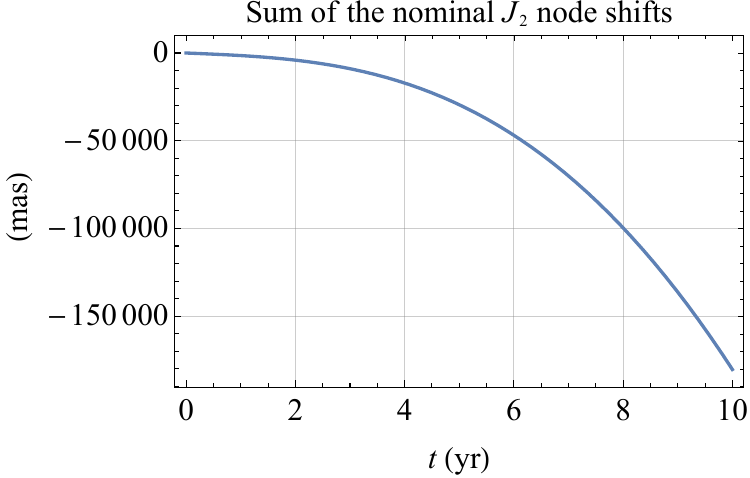}\\
\includegraphics[width = 8 cm]{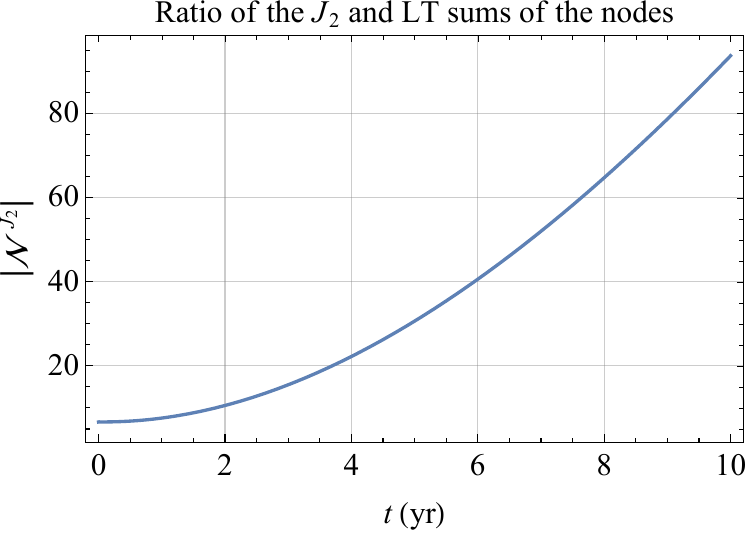}\\
\end{tabular}
\caption{Sums of the nominal  LT (upper panel) and $J_2$ (middle panel) shifts of the nodes, in mas, of a pair of counter--orbiting satellites numerically integrated over 10 years. The temporal variations of both $\boldsymbol{\hat{k}}$ \cite[pp.\,176-177]{2000Monte} and $J_2$, modeled according to ITSG-Grace2018, were included as well. An initial epoch 35 years after J2000.0 was assumed. An orbital height of $2\,000$ km was adopted for both satellites up to an offset of 10 km. The initial values of the inclinations differ from \rfrs{IAB}{OAB} and \rfrs{pol1}{pol2} by 10 mas. The lower panel shows the plot of the absolute value of \rfr{DeltaOJ2LT}.
}\label{figure2}
\end{figure}

\lb{sub:nodi}
By defining
\eqi
\mathcal{N}^{J_2}\ton{t}:=\rp{\Delta \Omega^{J_2}_\mathrm{A}\ton{t} + \Delta \Omega^{J_2}_\mathrm{B}\ton{t}}{\Delta \Omega^\mathrm{LT}_\mathrm{A}\ton{t} + \Delta \Omega^\mathrm{LT}_\mathrm{B}\ton{t}}\lb{DeltaOJ2LT},
\eqf
it is possible to repeat the previous analysis also for the sum of the nodes of POLARES. The results are in Figure \ref{figure2}.
The combined LT signature reaches the arcsec level over 10 year, while the nominal bias due to $J_2$ is up to $\simeq 90$ times larger than the former over the same time span.
\section{The LAGEOS-LARES 2 case}\lb{cazzate}
Recently, Ciufolini and coworkers \cite{Ciufoepjc24} claimed that LAGEOS and LARES 2 will allow them to perform a test of the LT effect accurate to $\simeq 0.2\%$  by monitoring the sum of their nodes, in accordance with the earlier proposal put forth by Ciufolini in \cite{1986PhRvL..56..278C}.

\begin{figure}
\centering
\begin{tabular}{c}
\includegraphics[width = 8 cm]{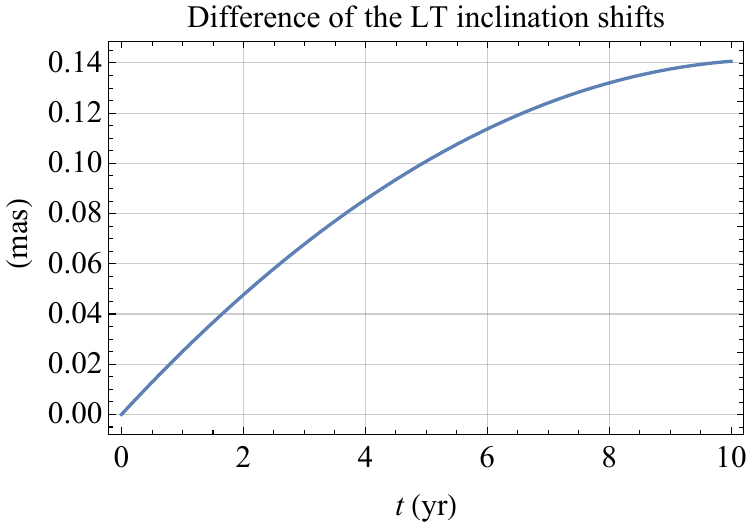}\\
\includegraphics[width = 8 cm]{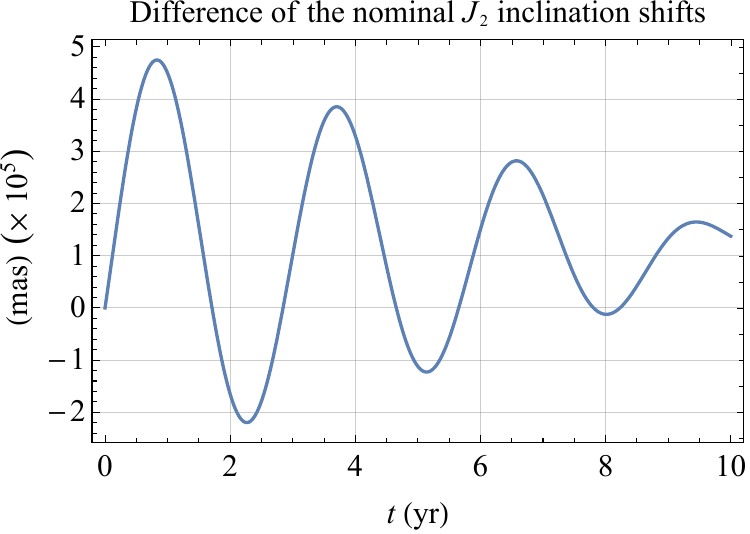}\\
\includegraphics[width = 8 cm]{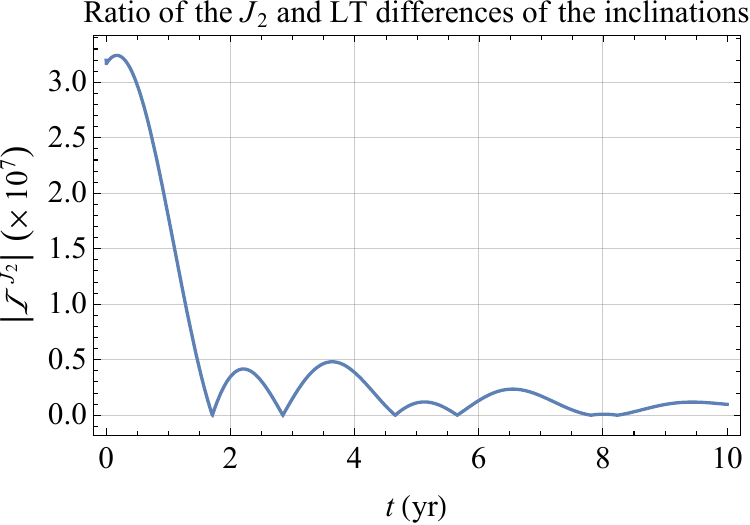}\\
\end{tabular}
\caption{Differences of the nominal  LT (upper panel) and $J_2$ (middle panel) shifts of the inclinations, in mas, of LAGEOS and LARES 2 numerically integrated over 10 years. The temporal variations of both $\boldsymbol{\hat{k}}$ \cite[pp.\,176-177]{2000Monte} and $J_2$, modeled according to ITSG-Grace2018, were included as well. The launch date of  LARES 2 was assumed as initial epoch. The initial values of the satellites' semimajor axis, eccentricity and inclination were retrieved from \cite[Tab.\,1]{2023EPJC...83...87C}, while those of the nodes were calculated with the WEB resource \url{https://www.n2yo.com/}. The lower panel shows the plot of the absolute value of \rfr{DeltaIJ2LT}.
}\label{figure3}
\end{figure}

\begin{figure}
\centering
\begin{tabular}{c}
\includegraphics[width = 8 cm]{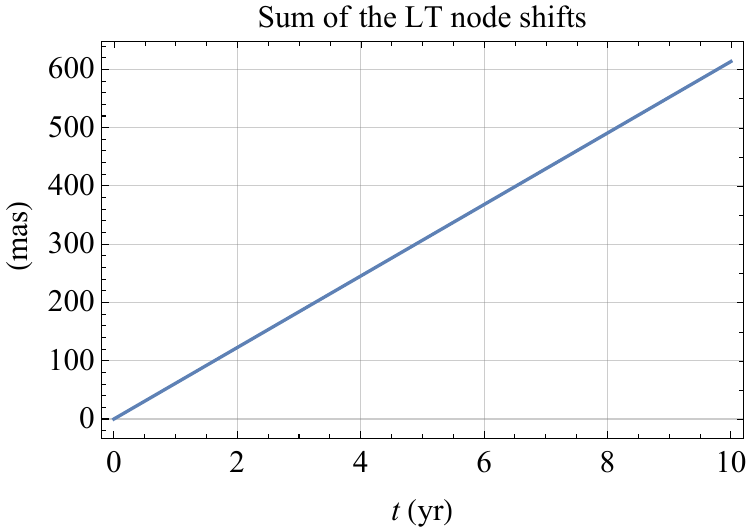}\\
\includegraphics[width = 8 cm]{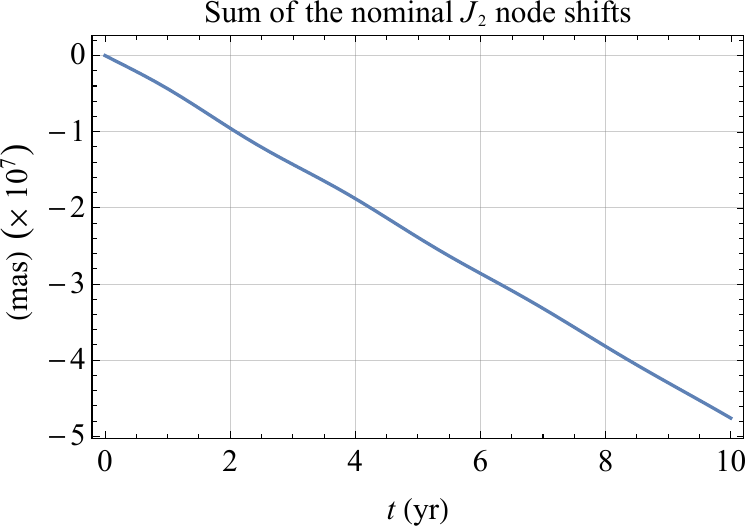}\\
\includegraphics[width = 8 cm]{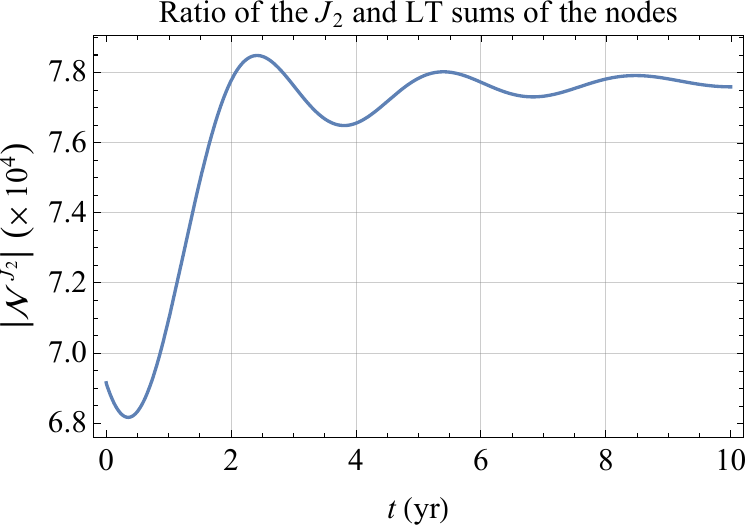}\\
\end{tabular}
\caption{Sums of the nominal  LT (upper panel) and $J_2$ (middle panel) shifts of the nodes, in mas, of LAGEOS and LARES 2 numerically integrated over 10 years. The temporal variations of both $\boldsymbol{\hat{k}}$ \cite[pp.\,176-177]{2000Monte} and $J_2$, modeled according to ITSG-Grace2018, were included as well. The launch date of  LARES 2 was assumed as initial epoch.  The initial values of the satellites' semimajor axis, eccentricity and inclination were retrieved from \cite[Tab.\,1]{2023EPJC...83...87C}, while those of the nodes were calculated with the WEB resource \url{https://www.n2yo.com/}. The lower panel shows the plot of the absolute value of \rfr{DeltaOJ2LT}.
}\label{figure4}
\end{figure}

In fact, all of the analyses by Ciufolini and coworkers over the years has always been based on \rfr{d90} and \rfrs{OLT}{OJ2}, which were not satisfied since the very epoch of the LARES 2 launch, occurred about 22 years after J2000.0. On the other hand, even if \rfrs{OLT}{OJ2} could be applied to the LAGEOS--LARES 2 experiment, the present author showed in \cite{2023Univ....9..211I} that the ambitious goal by Ciufolini and coworkers \cite{Ciufoepjc24} could not be met because of the consequences of the imperfect cancellation of the summed $J_2$--driven node precessions. By repeating the same analysis as in Section \ref{orb_err}, one gets Figures \ref{figure3} to \ref{figure4} which clearly exemplify how it is not possible to achieve the accuracy goal stated in \cite{Ciufoepjc24}, not even looking at the difference of the inclinations.
%
%
%
%
%
%
%
%
\section{Summary and conclusions}\lb{fine}
It has been shown that, for a general orientation of the primary's spin axis with respect to the inertial reference frame adopted, the Lense--Thirring rates of change of the orbital inclinations of two satellites moving along ideally identical orbits in opposite directions are equal and opposite, while those induced by the primary's oblateness have the same sign. However, the opposite happens for the nodes: the relativistic rates are the same, while the classical ones differ by a minus sign.  Thus, the difference of the inclinations and the sum of the nodes of two counter--revolving spacecraft allow, in principle, to cancel out the aliasing Newtonian shifts due to the quadrupole mass moment of the central body and enhance the gravitomagnetic ones.

The earlier proposal by Van Patten and Everitt--here branded POLARES--of using a pair of drag--free spacecraft moving in opposite directions along identical circular orbits passing through the Earth's poles is reexamined in view of the aforementioned results. Indeed, they would be applicable to
a hypothesized new mission since at the time of its launch, still to come, the terrestrial spin axis would be displaced with respect to its orientation at the epoch J2000.0, which is substantially assumed as reference $z$ axis of the geocentric inertial reference system usually adopted in actual satellites' data reductions.

For an orbital altitude of, say, $2\,000$ km, the combined relativistic inclination and node shifts of POLARES would amount to a few milliarcseconds and a couple of arcseconds, respectively, after 10 years from the launch.
By assuming not too stringent orbital injection errors, the nominal ratios of the signatures due to the Earth's first even zonal harmonic to the Lense--Thirring ones in both the differences of the inclinations and the sum of the nodes can be kept to a level sufficiently low to allow the indirect consequences on them of errors on the various physical and orbital parameters to be considered negligible.

It is not the case of the ongoing LAGEOS--LARES 2 experiment, both for the sum of the nodes and the difference of the inclinations, because of the imperfect cancellation of the classical orbital shifts for both the orbital elements.

Should the POLARES concept be implemented with passive, geodetic satellites of LAGEOS--type, a detailed investigation of several non--gravitational accelerations affecting them would be needed: it is outside the scopes of the present paper.
\section*{Data availability}
No new data were generated or analysed in support of this research.
\section*{Conflict of interest statement}
I declare no conflicts of interest.
\bibliography{Megabib}{}
\end{document}